# Insights into the Structural and Orientational Ordering of 2D Metal Halide Perovskites

Mustafa Mahmoud Aboulsaad[a,*], German Alvarez Carrero[a], Trupthi Devaiah Chonamada[b], Germán Salazar Alvarez[a], Ute Cappel[b,c], Rafael B. Araujo[a], Tomas Edvinsson[a,*]

[a]Department of Materials Science and Engineering, Solid State Physics, Uppsala University, Box 35, 75103 Uppsala, Sweden

[b]Department of Physics and Astronomy, X-ray Photon Science, Uppsala University, Box 516, 75120 Uppsala, Sweden

[c]Wallenberg Initiative Materials Science for Sustainability, Department of Physics and Astronomy, Uppsala University, 75120 Uppsala, Sweden

[*]Corresponding authors: mustafa.aboulsaad@angstrom.uu.se, tomas.edvinsson@angstrom.uu.se

## Abstract

Metal halide perovskite nanoplatelets combine quantum confinement with structural anisotropy, yet how thickness and processing govern their internal crystal lattice versus collective organization remains unresolved. In this study, we separate internal structure, superlattice organization, and film texture using 2-5 monolayer $CsPbBr_3$ nanoplatelets and larger nanocrystals across spin-coating and drop-casting routes. Grazing incidence X-ray diffraction analysis of the specimens indicates that orthorhombic-compatible signatures recur throughout the thickness and deposition series, with a preserved lattice–morphology relationship in which 020/040 and 101/202 lattice planes are associated with the platelet thickness and lateral dimensions, respectively. Upon deposition on the substrate, we notice a thickness-dependent crossover: 2-monolayer nanoplatelets remain predominantly face-on across deposition routes, whereas thicker nanoplatelets show marked processing sensitivity and orientational disorder. Processing therefore can be used as a tool to direct the NPL packing, establishing thickness as a key determinant of processing sensitivity.

**Keywords:** perovskite, nanoplatelets, GID, orientation, structure

## Main

Colloidal metal–halide perovskite nanoplatelets (NPLs) have emerged as a versatile platform for studying quantum confinement and anisotropic optoelectronic properties in low-dimensional semiconductors. In all-inorganic $CsPbBr_3$, solution syntheses can now deliver platelets whose thickness is defined in integer numbers of $PbBr_6$ octahedral layers, enabling monolayer-level control of bandgap and exciton binding energy. As a result, $CsPbBr_3$ NPLs exhibit narrow emission bands, high radiative recombination fractions, and tunable photoluminescence across the blue–green spectral range simply by varying the number of monolayers. [1–4] At the same time, these strongly confined nanocrystals are structurally anisotropic and readily self-assemble into stacked films or superlattices, so that their solid-state phase, lattice distortion, and orientation distribution cannot be inferred directly from bulk $CsPbBr_3$.[3,5,6] This coupling between internal structure and collective organization makes it difficult to distinguish genuine crystallographic evolution from changes in diffraction visibility caused by finite-size broadening, preferred orientation, and superlattice formation.

Recent structural studies have begun to resolve how reducing the thickness of $CsPbBr_3$ to a few monolayers (ML) perturbs its crystal structure. Bertolotti et al. combined X-ray total scattering and electron microscopy to show that 6 ML-thick $CsPbBr_3$ NPLs (1 ML ≈ 0.59 nm) adopt an orthorhombic perovskite structure (*Pnma*) at room temperature, with the platelet surfaces parallel to {101} planes and with measurable anisotropic lattice strain relative to the bulk[5]. Using multilayer diffraction on highly ordered stacks of colloidal platelets, Toso et al. refined the structure of ~12 Å-thick Cs–Pb–Br NPLs, determining their thickness, stoichiometry, surface passivation, and small but significant deviations of lattice parameters from bulk $CsPbBr_3$[6]. When viewed together with thickness-controlled synthesis and optical data, these measurements indicate that decreasing the number of perovskite monolayers progressively modifies both lattice metrics and effective phase stability, with ultrathin platelets deviating more strongly from the bulk orthorhombic structure while retaining the corner-sharing perovskite framework[7]. In a recent study, we proposed that the coexistence of cubic and orthorhombic phases at room temperature in 2 ML $CsPbBr_3$ NPLs is energetically allowed, but as the thickness increases, the orthorhombic phase is successively stabilized[8].

In a conventional XRD study, Wang et al.[9] showed that increasing the number of monolayers leads to a phase evolution from cubic to orthorhombic. Resolving such thickness-dependent structural evolution, however, remains challenging when local structural information is obtained from FFT analysis of TEM images. Although the measured spacings can be indexed using the cubic phase, this assignment should be treated with caution because of the strong structural similarity between cubic and orthorhombic $CsPbBr_3$. For example, the orthorhombic *Pnma* structure has similar d-spacings between the 202 and 040 planes, and also between the 101 and 020 planes, which can be mistaken for the 200 and 100 planes of the cubic structure[10,11]. Therefore, HR-TEM and FFT analysis provide important local structural information, but they cannot always be used alone to unambiguously identify the crystal polymorph in $CsPbBr_3$ NPLs[1,4,12–14]. This is further complicated by the beam sensitivity of these materials, since electron-beam exposure can degrade the phase structure and induce Pb displacement, leading to an additionally distorted structure[8,15]. Similar coexistence of both structures within the same ensemble has also been reported in other studies, further supporting the idea that local structural heterogeneity can exist even when the ensemble scattering response is dominated by orthorhombic features[16–18]. Together, these observations highlight the importance of combining local imaging with ensemble diffraction when evaluating the structural evolution of thickness-controlled NPLs.

Beyond this structural thickness dependence, thin-film formation introduces a second level of hierarchical complexity. During in situ-grown $CsPbBr_3$ quantum-well thin films, the chemical conditions can modify crystallization kinetics and the resulting distribution of NPL thicknesses[19]. In contrast, pre-formed colloidal perovskite NPLs can reorganize during solvent evaporation to form ordered multilayer stacks whose periodicity and structural parameters depend on platelet thickness and ligand-mediated surface passivation[6]. More broadly, oriented halide perovskite nanostructure thin films obtained by colloidal processing can be engineered to favor either “face-on” or “edge-on” alignment of their inorganic layers, with substantial consequences for charge transport and light extraction[20]. In situ-grown perovskite NPLs emission layers comprising a monolayer of face-on nanoplatelets already achieve high fractions of horizontally oriented transition dipole moments and high external quantum efficiencies in planar LEDs, where solvent-controlled self-assembly of $CsPbI_3$ NPLs superlattices enables either face-on or edge-on platelet orientation and highly linearly polarized electroluminescence[21,22].

Several reports have therefore considered the effect of deposition method on the orientation of synthesized colloidal nanoplatelets, but direct comparison remains difficult because the studied materials have different monolayer thicknesses, solvents, and thin-film formation pathways. Ye et al.[22] showed that fast and slow annealing of spin-coated $CsPbI_3$ 3 ML-NPLs samples yields edge-on or face-on NPLs, respectively. Krajewska et al.[23] reported the opposite trend for $CsPbBr_3$ NPLs, where face-on and edge-on NPLs resulted from fast and slow evaporation, respectively, but in drop-cast samples of 2 ML NPLs only. More recently, Lee et al. showed that ligand engineering can tune self-assembled $CsPbBr_3$ NPL superlattices from face-on to edge-on orientations by modifying both inter-ligand interactions and NPL lateral dimensions, demonstrating that orientation is also strongly coupled to surface chemistry and particle geometry.[24] Taken together, these studies show that nanoplatelet orientation is dependent on the number of MLs, lateral dimensions, ligand chemistry, and film processing conditions.

In this work, we address three closely related questions concerning the structural and orientational behavior of $CsPbBr_3$ nanoplatelets: whether their crystal structure evolves with the number of monolayers, how the number of monolayers affects their preferred orientation, and how the deposition/drying pathway influences their thin-film organization and apparent growth direction. We present a systematic investigation in which monolayer thickness and deposition pathway are compared, whereas the solvent and ligand system are maintained within the same experimental framework.

To this end, we synthesized $CsPbBr_3$ nanoplatelets with thicknesses from 2 to 5 ML, together with larger nanocrystals (edge lengths between 7.6 ± 2.0 and 12.5 ± 2.5 nm) for comparison, and investigated the as-deposited thin films using grazing incidence X-ray diffraction (GID). For each thickness, thin films were prepared using spin coating (Spin), slow solvent-evaporation drop-casting (DS), and fast solvent-evaporation drop-casting (DF). Importantly, all thin films were deposited from the same solvent, hexane, allowing the effect of solvent polarity to be excluded and enabling a direct comparison of drying kinetics and deposition method. In $CsPbBr_3$ nanoplatelet thin films, monolayer thickness and deposition pathway mainly control thin-film orientation, superlattice coherence, and interparticle organization, whereas the internal orthorhombic-compatible structural motif and lattice–morphology relationship remain largely preserved across 2–5 ML nanoplatelets. We therefore present a systematic GID-based framework that separates structural assignment from texture/orientation effects and clarifies how thickness and processing govern the organization of colloidal halide perovskite nanoplatelet thin films. By varying NPL thickness and deposition pathway within the same material and solvent system, this comparison determines whether deposition-induced orientation trends are universal or thickness dependent. The results reveal that they are thickness dependent: 2 ML NPLs retain predominantly face-on alignment across all three deposition routes, whereas thicker NPLs exhibit greater sensitivity to drying kinetics. Consequently, orientation trends established for one NPL thickness should not be generalized to another, and internal lattice structure, the crystallographic lattice–morphology motif, local superlattice organization, and global film texture must be evaluated as distinct levels of structural organization.

## Halide perovskite nanoplatelets: orientation and GID

Unlike conventional polycrystalline perovskite thin films, colloidal nanoplatelet thin films are assembled from preformed, single-crystalline platelets, and their final organization is strongly influenced by interparticle interactions and the deposition process[3,25]. Their GID patterns contain information about both the internal crystal structure of the individual nanoplatelets and their collective orientation within the thin film. Peak positions and peak splitting primarily provide information about lattice symmetry and crystal phase, whereas the intensity distribution in reciprocal space reflects thin-film texture and platelet orientation.[26] Here, face-on NPLs are defined as platelets with their large basal faces (0*k*0) lying flat on the substrate, whereas edge-on NPLs have their basal faces approximately perpendicular to the substrate. We apply this analysis to thin films prepared by spin coating and drop-casting, as illustrated schematically in Fig. 1a.

The optical spectra provide the first validation of the thickness-controlled NPLs series. The photoluminescence (PL) spectra in Fig. 1b exhibit narrow peaks, with room-temperature FWHM values ranging from 0.1 eV for 2 ML NPLs to 0.17 eV for 5 ML NPLs, and the emission shifts from approximately 2.85 eV for 2 ML NPLs to 2.54 eV for 5 ML NPLs as the thickness increases. This thickness-dependent red-shift is characteristic of nanoplatelets because quantum confinement weakens as the NPLs thickness increases. The larger nanocrystals emit shifted further to the red, at approximately 2.43 eV with FWHM $\approx$ 0.11 eV, consistent with weak quantum confinement in large NCs[27,28]. This trend aligns well with reported PL spectra in the literature, as well as our own previous findings[1,8,29,30].

The deposition pathway also affects the optical response without indicating a separate crystallographic phase change. Both the DF and DS spectra exhibited typical homogeneous peak broadening, consistent with our earlier observations[8]. For the 2 ML samples, a minor PL peak shift was observed between the DS and DF preparation methods; this discrepancy diminished with increasing monolayer thickness and was minimally present for the NC1 sample, where NC1 corresponds to the larger and more shape-homogeneous nanocrystals obtained by the highest Cs/Pb ratio during the synthesis. In the DF configuration of the NC2 sample, where NC2 is the larger and shape-homogeneous nanocrystals obtained as a byproduct of the first purification step of NC1, the PL emission split into three major peaks, indicating the formation of mixed-size domains ranging from small nanoclusters of 3–4 ML thickness to larger aggregates that mimic the PL response of bulk $CsPbBr_3$ (Fig. S16d). This is a typical consequence of rapid solvent evaporation, which alters not only structural arrangement but also thin-film homogeneity through the formation of island-like coatings[31,32]. Conversely, the Spin 5 ML NPLs sample showed an asymmetric peak shoulder, hinting at the formation of alternative thicknesses. Because this deviation was observed for only one sample, whereas the preceding and subsequent Spin samples followed the expected trend, we consider it an isolated sample-specific variation rather than a systematic effect of the spin-coating process. The origin of this deviation cannot be determined conclusively and may reflect experimental or data-related variability. Across all studied systems, the Spin samples displayed a slight blue-shift relative to their DF and DS counterparts, regardless of whether the system was strongly or weakly confined. This difference is consistent with stronger particle–particle interactions during the slower solvent evaporation in DF and DS casting than during spin coating. Such interactions can facilitate exciton transfer and modify excitonic coupling, providing a plausible origin for the observed red-shift in PL emission, as reported in related systems[33]. Such interparticle interactions are expected to be reduced in the Spin thin films due to rapid solvent evaporation, which curtails the interaction window even when compared to drop-cast thin films on substrates. Although these optical variations provide information about thin-film homogeneity and interparticle coupling, they cannot by themselves distinguish changes in particle organization from changes in crystal structure. We therefore turn to GID analysis, for which a careful distinction between detector-space and reciprocal-space coordinates is essential when assigning nanoplatelet orientation.

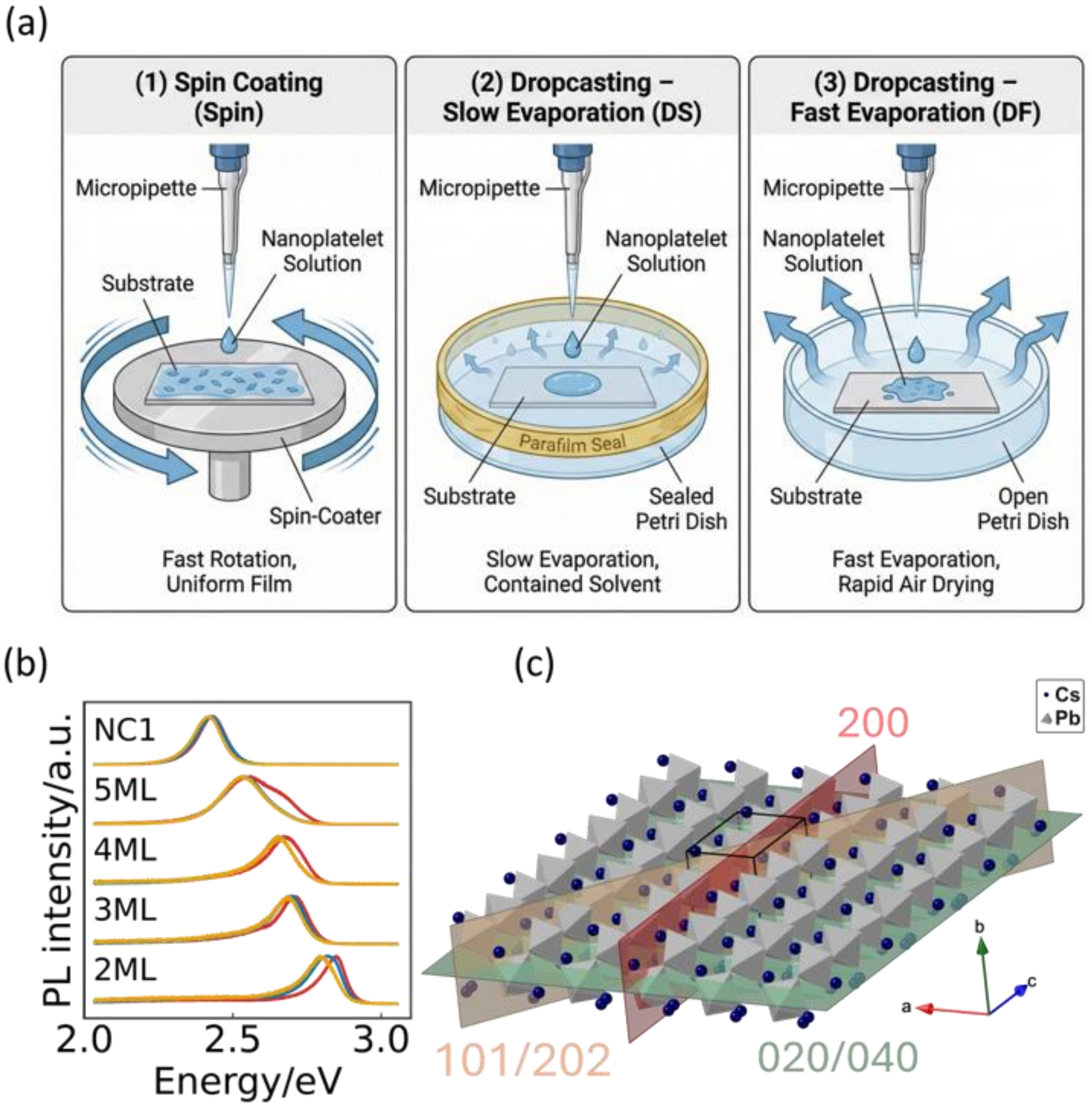

**Figure 1:** (a) Illustrative diagram showing the different deposition strategies followed in this work. (b) The normalized PL spectra in a PL-Raman confocal system of NC1 and 2-5 ML NPLs thin films for each deposition strategy. Blue, orange, and red represent DF, DS, and Spin, respectively. (c) Structural model of orthorhombic $CsPbBr_3$ showing the 200, 020/040, and 101/202 crystallographic planes used for peak assignment and orientation analysis. The blue spheres and grey octahedra correspond to Cs ions and $PbBr_6$ units, respectively.

## Growth and structural analysis

We first consider the structural assignment and the lattice–morphology relationship of the colloidal NPLs. FFT analysis was performed on DF-2 ML, DF-5 ML, DS-5 ML NPLs, as well as on DF-NC1 and DS-NC1 (Figs. S1h, S10h, S11g, S13g, and S14g). The analysis showed reflections with spacings of 0.59, 0.43, 0.30, and 0.27 nm for the 2 ML NPLs, and 0.59, 0.43, 0.30, and 0.27 nm for the 5 ML NPLs and NC1. These values correspond well to the 100, 110, 200, and 210 planes of the cubic phase. However, this cubic indexing should be treated with caution because cubic and/or orthorhombic $CsPbBr_3$ have strongly overlapping d-spacings for some essential planes in HRTEM[11]. As a reference, we compared our data with the bulk-like diffraction pattern of NC2, and *Pnma* from ICSD (205061). This sample clearly shows the presence of the orthorhombic phase (Fig. S16a-c)[10]. In the DF-4 ML NPLs, the splitting between the 040 and 202 reflections near $q \approx 2.15$ Å$^{-1}$ is directly visible in both the radially integrated profile and the azimuthally resolved scattering map (Figs. 2a,b and S17a,b), providing a clear orthorhombic-compatible signature.

For the DF-3 and DF-5 ML NPLs, the radially integrated profiles in Fig. 2a show asymmetric broadening near $q \approx 2.15$ Å$^{-1}$. To assess whether these profiles are compatible with the unresolved counterparts of the reflections observed directly in the intermediate DF-4 ML sample, the DF-3 and DF-5 ML profiles were described using three-component peak models in FitED[34]. The fitted component positions lie near the expected positions of the orthorhombic 040, 202, and 212 reflections (Figs. 2c and S10c). However, profile decomposition is not unique, because asymmetric broadening may also arise from strain-related lattice-parameter distributions, size and coherence effects, structural heterogeneity, texture-dependent broadening, or collective scattering, as in the case of the very thin NPLs[35]. The fitting is therefore treated as phenomenological supporting evidence alongside the directly resolved 040/202 splitting in DF-4 ML and the additional weak orthorhombic-compatible reflections, rather than as independent identification of the individual fitted components. The DF-2 ML case requires additional caution because its pronounced superlattice coherence may influence the width and shape of the feature near $q \approx 2.13$-$2.15$ Å$^{-1}$. Although a three-component model provides a satisfactory phenomenological description of the asymmetric profile (Fig. S1c), the components cannot be assigned uniquely to the 040, 202, and 212 reflections. The fitting is therefore not used as independent evidence for orthorhombic splitting in the 2 ML sample. Instead, the structural interpretation relies primarily on the weak orthorhombic-compatible features observed near 1.58, 1.68, and 2.0 Å$^{-1}$, together with the broader consistency of the complete GID dataset. Importantly, the same superlattice coherence that complicates interpretation of the 2.13-2.15 Å$^{-1}$ profile enables the periodic stacking analysis presented in the orientational section. After performing the same level of analysis and tracking the 2D GID maps, we observed that orthorhombic structural features exist in all monolayers with no clear trend favoring the orthorhombic structure versus the deposition method. In Table 1, we show which orthorhombic features exist in each monolayer system for each deposition strategy. The complete sample-specific 2D GID maps, azimuthal maps, radial profiles, low-$q$ profiles, and $q_z$-dependent integrations for the 2–5 ML NPLs and NC1 under all three deposition conditions are provided in Figs. S1–S15.

**Table 1:** Summary of the orthorhombic-compatible scattering signatures most clearly observed in the investigated NPL and NC1 systems. Direct peak splitting refers to visibly resolved components near $q \approx 2.15$ Å$^{-1}$, whereas multi-component fitting describes asymmetric profiles that are consistent with unresolved orthorhombic reflection overlap.

| Signature | Samples displaying the signature |
|---|---|
| Small features (1.60 – 2.04 Å$^{-1}$) | **DF** [2 ML NPLs and NC1]<br>**Spin** [3, 4 ML NPLs] |
| Peak splitting (2.15 Å$^{-1}$) | **DF** [4 ML NPLs]<br>**DS** [2, 3, 4 ML NPLs] |
| Asymmetric profile near 2.15 Å$^{-1}$ [multi-component fit] | **DF** [3, 5 ML NPLs] |

Complementary χ-sector and reciprocal-space α-sector integrations of the DF and DS series reveal systematic orientation-dependent shifts of the 020/101 and 040/202 features, providing convergent diffraction evidence that strongly supports a *Pnma*-compatible orthorhombic assignment and the 020/040–101/202 lattice–morphology relationship (Fig. 2d and Figs. S20–S22). As shown in Figs. 2b,e, the peaks at 1.07 and 2.15 $Å^{-1}$ become sharper in the in-plane $q_r$ direction, while the same peaks broaden in the out-of-plane $q_z$ direction. This relationship is illustrated more explicitly by the azimuthally resolved DF-4 ML data in Fig. S17b. In the $I(|q|, \chi)$ representation, intensity at low $\chi$ corresponds predominantly to in-plane scattering-vector components, whereas intensity near $\chi \approx 90°$ corresponds predominantly to out-of-plane components. The complementary angular distributions of the overlapping 101/020 reflections near 1.07 $Å^{-1}$ and their higher-order 202/040 counterparts near 2.15 $Å^{-1}$ therefore associate the 101/202 family with lateral crystalline coherence and the 020/040 family with the platelet-normal, thickness-related direction. In addition, the 200 and 121 contributions near 1.51 $Å^{-1}$ appear at approximately $\chi$ = 3° and 45°, respectively, but correspond to equivalent angular relationships with respect to $q_r$ after reciprocal-space remapping in this geometry (Figs. 2b and S17b). This illustrates why detector-space χ-values cannot be converted directly into sample-orientation angles. The same behavior is observed for all monolayer thicknesses, with the 121/200 feature becoming more ring-like as the number of monolayers increases. This provides clear evidence of in-plane anisotropy. In both NC1 and NC2, the contributions from all peaks become more equivalent in both $q_r$ and $q_z$, except for the 101 peak in the NC1 case. This arises from the higher symmetry of the NCs shape compared to that of the NPLs (see HR-TEM analysis in the next section). Similar behavior was observed for DS and Spin systems, with no change in the growth direction. However, in the Spin case, the contribution of the 040/202 planes in $q_z$ is weaker than their contribution in $q_r$ compared with the DS and DF cases, yet they are still broader, consistent with the same trend. These angular distributions are illustrated in the azimuthal intensity map in Fig. 2e. The observation that all $CsPbBr_3$ nanoplatelets from 2 to 5 ML, together with the more isotropic nanocrystals, show orthorhombic structural features regardless of the deposition strategy strongly suggests that the crystal structure is mainly established during colloidal growth and remains largely unaffected by the subsequent thin-film formation process. This conclusion is consistent with recent studies showing that the room-temperature structures of $CsPbBr_3$ nanocrystals and nanoplatelets are more accurately described by the orthorhombic phase than by a truly cubic lattice, even when electron microscopy images appear pseudocubic because of the small magnitude of octahedral tilting and the limited sensitivity of HRTEM to distinguish between closely related perovskite polymorphs[5,11]. In perovskites, the transition between cubic and orthorhombic structures is not primarily expressed as a simple stretching of the lattice parameters, but rather through distortion and tilting of the octahedral units, which can be especially difficult to resolve in very thin nanocrystals using local imaging alone[36]. In this context, the present GID results should be understood as an ensemble structural probe, whereas HR-TEM/FFT provides local but less definitive polymorph information.

A more reliable diffraction-based criterion for distinguishing cubic and orthorhombic $CsPbBr_3$ is the higher density of diffraction peaks expected for the orthorhombic phase compared with the cubic phase. In particular, two characteristic features can be used to distinguish *Pm-3m* from different orthorhombic space groups: the presence of several weak reflections between 1.60 and 2.04 $Å^{-1}$, and the splitting of the cubic 200-related peak around 2.15 $Å^{-1}$.[10] Due to preferred orientation, only one of these two features may be clearly visible in a given 2D scattering geometry, but the presence of either feature still supports assignment to the orthorhombic structure[10]. For this reason, robust GID interpretation should not rely solely on fully azimuthally integrated 1D profiles but should also consider the complete 2D reciprocal-space maps together with complementary azimuthal maps and $q_z$-dependent reciprocal-space integrations. Because scattering from textured crystallites may be confined to limited azimuthal regions, integration over the full azimuthal range can reduce the peak-to-background contrast of weak, orientation-specific reflections and consequently obscure subtle structural-phase signatures[26,37]. This becomes even more crucial for very thin nanocrystals, where Scherrer broadening, preferred orientation, and collective scattering from superlattice ordering can obscure weak peaks and small shoulders[35].

The 2 ML NPLs represent the most delicate case because the narrow scattering feature around 2.15 $Å^{-1}$ can be enhanced by collective scattering from the superlattice, whereas the associated peak asymmetry can have more than one origin. It may reflect the reflection splitting expected for the orthorhombic structure; however, in the two-octahedral-layer-thick NPLs, it may also be influenced by surface- and ligand-induced lattice distortions, whose contribution is amplified by the exceptionally high surface-to-volume ratio, or by ligand-mediated restriction of octahedral tilting, as demonstrated in related all-inorganic halide-perovskite systems[6,38]. Nevertheless, the additional weak reflections observed in the 1.60–2.04 Å$^{-1}$ region support the presence of orthorhombic structural contributions even in the thinnest NPLs. The multi-component fit to the 2.15 Å$^{-1}$ profile is compatible with this interpretation but, for the reasons discussed above, is not treated as independent structural evidence[10,35]. The absence of any detectable deposition-induced phase change, within the sensitivity of the present GID measurements, therefore, indicates that varying the drying rate or deposition route primarily modifies texture and orientational order rather than the internal crystallographic structure.

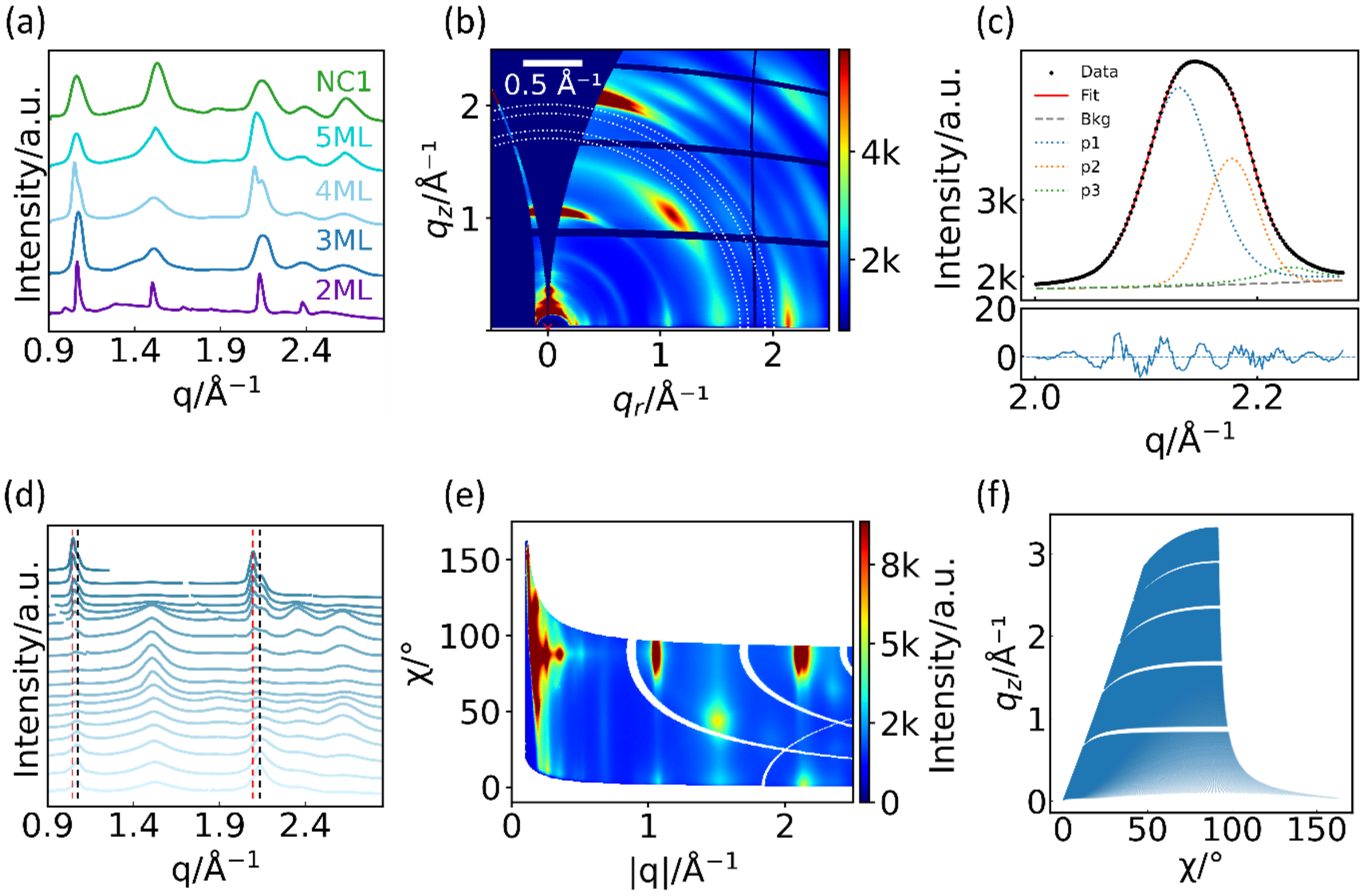

**Figure 2:** (a) Vertically offset 1D GID profiles of the 2–5 ML NPLs and NC1 of DF samples, showing the thickness-dependent evolution of the scattering features. (b) Representative 2D GID pattern displayed in ($q_r$,$q_z$) coordinates at an incidence angle of 0.4° for DF-4 ML NPLs; the dotted rings mark selected scattering-vector positions of the small orthorhombic features. (c) Three-component fit to the asymmetric scattering profile near $q \approx 2.15$ Å$^{-1}$ for DF-3 ML NPLs, shown together with the experimental data, total fit, background, individual components p1–p3, and fitting residual. The fitted component positions lie near the expected positions of the orthorhombic 040, 202, and 212 reflections, but the fit is presented as a phenomenological description rather than a unique separation of these reflections. (d) 1D GID intensity profiles of DF-4 ML NPLs thin film. The profiles were obtained by angular integration in reciprocal space using $\alpha = \tan^{-1}(q_z/q_r)$. Profiles were extracted from angular sectors centred from 5° to 90° in 5° increments, with a half-width of 2.5°, and are vertically offset for clarity from low to high angle. The red and black dashed lines indicate the peak-component centres of the local maxima at high angles (out-of-plane) and low angles (in-plane), respectively. (e) Azimuthal intensity map, $I(|q|, \chi)$, illustrating the angular distribution of the principal reflections and the preferred crystallographic orientation for DF-4 ML NPLs. (f) General representation of the accessible azimuthal angle in detector space ($\chi$) and the out-of-plane component of the scattering vector in reciprocal space ($q_z$) from our specific configuration, after accounting for the missing wedge and how $\chi$ and $q_z$ relate to each other.

Across all thicknesses, the GID analysis reveals a preserved lattice–morphology relationship, with thickness-related coherence associated predominantly with the 020/040 reflections and lateral coherence with 101/202, indicating that this anisotropy is established during NPL synthesis and retained upon film formation rather than generated during deposition. This lattice–morphology relationship and the corresponding crystallographic planes are schematically illustrated in Fig. 1c. More broadly, the morphology of colloidal semiconductor nanoplatelets is known to emerge from anisotropic facet energetics, layer-addition thermodynamics, and surface chemistry, as demonstrated for II–VI NPL systems[39,40], suggesting that coupling between crystal structure and platelet geometry is a general feature of strongly anisotropic nanocrystals. Within $CsPbBr_3$ specifically, however, the crystallographic origin of this morphology has been described differently across the literature. Bertolotti et al. assigned 6-ML NPLs to *Pnma* symmetry with {101} basal facets and lateral extension involving {010} and orthogonal {101} facets[5], whereas again in another study Bertolotti et al. reported a distinct geometry for nanocrystals, with (001) as the basal plane and {110} associated with lateral growth[36]. By contrast, Zhu et al. identified *Pnma* $CsPbBr_3$ NPLs with 020/040 associated with the thickness direction and 101/202 with the lateral dimensions[11], in close agreement with the relationship observed here. Its persistence across the present thickness and processing series therefore supports an intrinsic lattice–morphology relationship in these $CsPbBr_3$ NPLs that remains largely unaffected by subsequent thin-film processing. Our results are in strong agreement with the crystallographic dimensional relationship reported by Zhu et al., where the 020/040 family is associated with the thin dimension and the 101/202 family with the lateral dimension[11]. The recurrence of the corresponding angular-intensity distributions across the complete sample set (Figs. S1–S15), with DF-4 ML presented as the explicitly indexed example in Fig. S17b, supports the persistence of this lattice–morphology relationship across NPL thicknesses and deposition pathways. The additional contribution of the 121/200 feature in the in-plane direction and around 45° relative to $q_r$ suggests a tilted side-plane contribution to the anisotropic growth geometry. The difference in $q_r$ peak broadening between our work[8] and Zhu et al. work[11] also suggests a lower degree of ordering in our 3-5 ML NPLs assemblies. In their case, the $q_r$ peaks are sharper, whereas in our samples, they are slightly broader, indicating less ordered nanoplatelet assemblies.

The same growth-direction relationship was observed for the DS, DF, and Spin systems, confirming that the deposition strategy does not change the internal growth motif of the NPLs. However, in the Spin thin films, the weaker $q_z$ contribution of the 040/202 planes can be understood as a consequence of thin-film morphology and orientation rather than a change in crystal structure. This weaker contribution arises from the intensity reduction due to larger broadening, fewer stacked NPLs on the substrate because of the centrifugal coating process, and/or the higher-order nature of the peak around 2.15 $Å^{-1}$, which makes it more sensitive to disorder in highly oriented samples.

Taken together, these results suggest that monolayer thickness primarily influences lattice relaxation, strain, scattering coherence, and the degree of orientational order, whereas the *Pnma*-compatible orthorhombic signatures and associated lattice–morphology relationship are robust features inherited from the colloidal nanoplatelets prior to deposition. Therefore, the previously proposed thickness-dependent cubic-to-orthorhombic transition should not be viewed as a simple conversion from cubic 2 ML NPLs to orthorhombic thicker NPLs, but rather as an evolution in the visibility, coherence, and dominance of orthorhombic scattering features within a structurally heterogeneous nanoplatelet ensemble. Having established that Pnma-compatible orthorhombic signatures and the lattice–morphology relationship persist across the thickness and deposition series, we now turn to what deposition does control. If the internal crystallographic framework is set during colloidal synthesis and is insensitive to thin-film formation, then any differences between samples prepared by spin coating, fast solvent-evaporation drop-casting, and slow solvent-evaporation drop-casting are therefore attributed primarily to differences in assembly, i.e., how the nanoplatelets orient relative to the substrate and to one another. The following section quantifies this orientational behavior systematically across the full thickness series, which also allows us to directly address the conflicting orientation trends reported in the literature for different nanoplatelet systems.

## Orientational analysis

For nanoplatelets to be effectively implemented in optoelectronic devices, efficient carrier transport is essential for achieving high performance. Bulk halide perovskites are well known for their excellent charge-transport properties[41]. However, in nanocrystal-based systems, this intrinsic advantage can be strongly influenced by additional factors, particularly the presence of insulating surface ligands and the orientation of the nanocrystals within the thin-film. For example, Chen et al.[42] demonstrated that controlling the orientation of perovskite nanoplatelets in photovoltaic devices can facilitate charge transport between the electrodes through the inorganic octahedral framework, thereby reducing the limiting influence of the organic ligands. Although this description may not fully capture the complexity of charge transport in such systems, where interparticle carrier transfer also plays an important role, it clearly highlights the critical importance of nanoplatelet orientation. Several studies have investigated nanoplatelet orientation and superlattice formation by comparing thin films prepared under different solvent-evaporation conditions, particularly slow and fast solvent evaporation during drop-casting[22,23]. These works have provided important insights into the relationship between drying conditions, self-assembly, and crystallite orientation. However, most of these studies focus on a particular nanoplatelet thickness, and the investigated monolayer number often differs from one study to another. This makes direct comparison between reported orientation trends challenging and may lead to misleading conclusions when the effect of nanoplatelet thickness is not considered explicitly. Here, we provide a systematic comparison of nanoplatelet orientation across a series of well-defined monolayer thicknesses, from 2 to 5 ML, and benchmark these results against larger nanocrystals. This approach enables the influence of nanoplatelet thickness and deposition method on crystallite orientation to be evaluated within the same experimental framework.

Nanoplatelet crystallite orientation was analyzed using the mosaicity factor (MF) approach[43]. For this analysis, the second superlattice reflection (002), located at approximately 0.25–0.28 $Å^{-1}$, was azimuthally integrated. The relative contributions from face-on, corresponding to the out-of-plane orientation, and edge-on, corresponding to the in-plane orientation, were then quantified using the analysis code by Ogle et al.[43]. Owing to geometrical limitations imposed by the detector configuration, only approximately half of the signal associated with the in-plane reflection could be accessed experimentally. Therefore, to provide a more representative estimate of the edge-on population, the percentage of edge-on nanoplatelets extracted from this analysis was corrected by a factor of two. The summarized orientation fractions as a function of monolayer number and deposition method are shown in Fig. 3g, whereas the corresponding alignment quality of the fitted face-on and edge-on populations, evaluated independently of their relative fractions, is presented in Fig. S18. The results show that nanoplatelet orientation is strongly governed by both the drying kinetics during thin-film formation and the nanoplatelet thickness/crystalline domain size. The drying rate follows the general trend: Spin > DF > DS. Under the fastest drying conditions, namely spin coating, the nanoplatelets preferentially adopt a face-on orientation for all monolayer thicknesses investigated. In contrast, for DS samples, the degree of preferential face-on orientation decreases more rapidly with increasing monolayer number. This suggests that slower solvent evaporation provides sufficient time for nanoplatelet reorganization during thin-film formation, resulting in a larger fraction of randomly oriented or edge-on-oriented particles. A clear dependence of the preferred orientation on monolayer number is also observed, especially in the 2 ML case (Figs. S1-S3). In general, 2 ML nanoplatelets exhibit a very high degree of face-on orientation, which appears relatively insensitive to changes in the deposition method. In other words, modifying the drying conditions does not significantly promote their reorientation toward the edge-on configuration. Particle-size and aspect-ratio analysis was performed using the TEM images shown in Figs. 3a–f, S1i, and S2g. For the 2 ML NPLs, the face-on population exhibited a mean lateral dimension of 27.9 ± 12.2 nm, while the edge-on population showed a thickness of 1.3 ± 0.2 nm and an accessible lateral dimension of 16.5 ± 2.9 nm. These dimensions correspond to lateral-to-thickness aspect ratios of 21.5 ± 10.0 and 12.7 ± 3.0 for the face-on and edge-on populations, respectively. For the 5 ML NPLs, a thickness of 3.1 ± 0.2 nm and an accessible lateral dimension of 17.2 ± 4.3 nm yielded an aspect ratio of 5.5 ± 1.5. The NC1 particles were considerably more isotropic: the most isotropic particles exhibited comparable dimensions averaging 9.5 ± 3.2 nm, whereas the less isotropic particles showed long and short dimensions of 12.5 ± 2.5 and 7.6 ± 2.0 nm, respectively, corresponding to an aspect ratio of 1.7 ± 0.5. These morphological trends are consistent with orientation analysis, with the strongly anisotropic 2 ML NPLs exhibiting the greatest preference for face-on alignment, whereas particles with progressively lower aspect ratios show a greater probability of adopting multiple orientations. These findings directly address the apparent conflict between the orientation trends reported by Ye et al.[22] and Krajewska et al.[23] in earlier studies. Rather than a genuine disagreement, the discrepancy reflects the strong thickness-dependence of orientational behavior that neither study could reveal, since each was restricted to a single monolayer number. For 2 ML nanoplatelets, the face-on configuration is dominant regardless of whether thin films are spin-coated or drop-casted, demonstrating that at this extreme aspect ratio, geometric anisotropy overwhelms kinetic control. For thicker NPLs (3–5 ML), however, the drying rate matters substantially: slower evaporation allows sufficient time for reorganization, yielding larger fractions of edge-on and unoriented populations. This interpretation is consistent with recent observations that increasing the lateral dimensions of perovskite NPLs promotes stronger face-down alignment in deposited films[44].

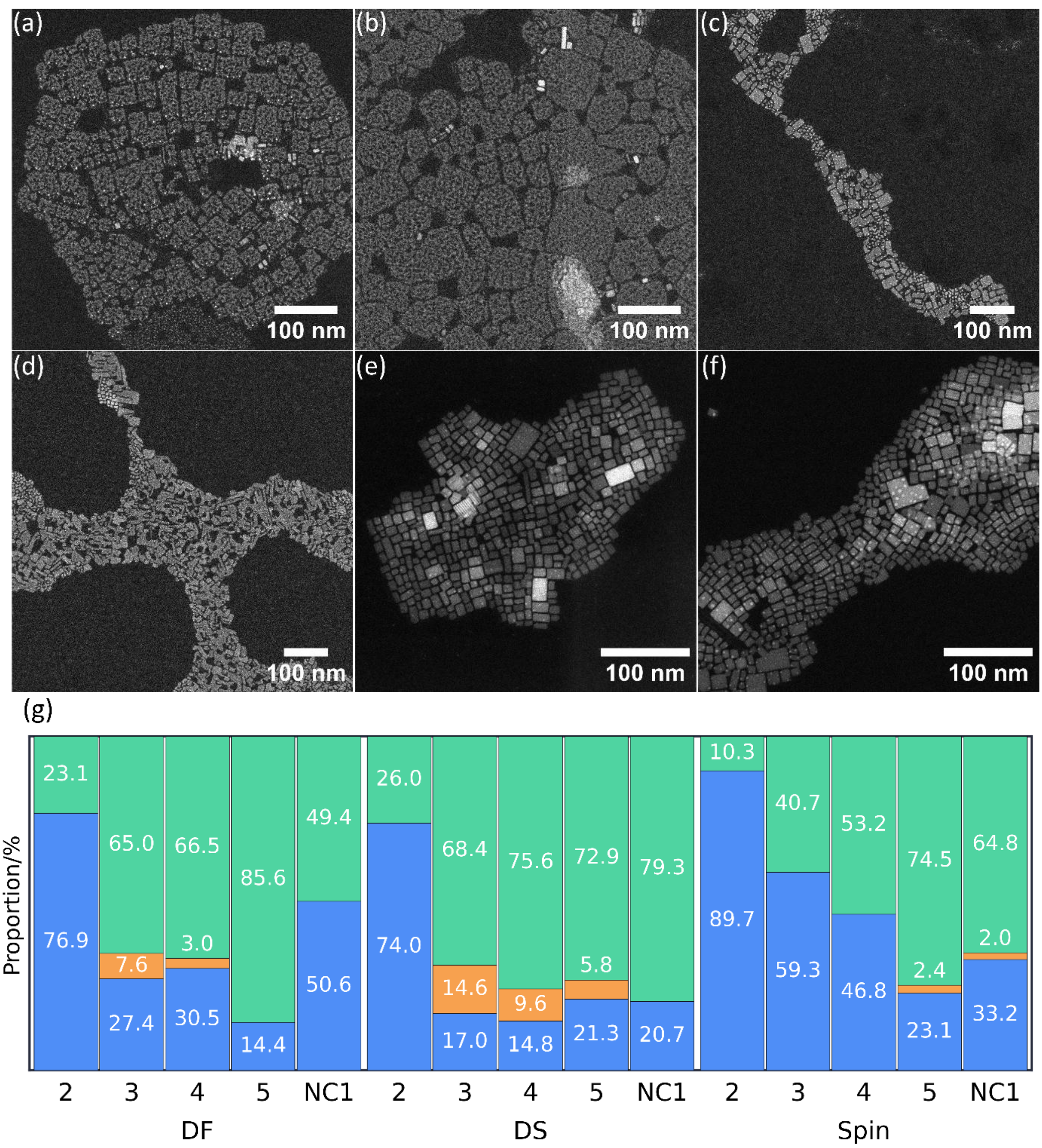

**Figure 3:** (a-f) TEM images of DF-2 ML, DS-2 ML, DF-5 ML, DS-5 ML, DF-NC1, and DS-NC1, respectively. (g) A normalized stacked bar plot of face-on (blue), edge-on (orange), and unoriented (green) orientation fractions as a function of monolayer number and deposition method. The numbers indicate the percentage of each orientation based on the MF approach.

The apparent conflict in literature therefore dissolves once thickness is treated as an independent variable rather than a fixed parameter. 3- and 4- ML NPLs, on the other hand, show a tendency to reorganize equally into the edge-on and face-on orientations, particularly under slow evaporation conditions, as observed for the DS samples (Fig. 3g), with the unoriented crystals still dominant. Overall, this analysis indicates that the NPLs with a number of MLs > 2 facilitate reorientation toward the edge-on configuration. This reorganization is further enhanced under slow evaporation conditions, such as those achieved in an evaporation-controlled chamber for DS nanoplatelet thin films. Figure 4a–c shows the $q_z$-dependent intensity profiles of the overlapping (121/200/002) reflection for the 2 ML DF, DS, and Spin thin films, respectively. The occurrence of this feature in the same $q_z$ region for all three preparation routes is consistent with the near-degenerate orthorhombic reflections expected for a *Pnma*-like lattice and indicates that the corresponding out-of-plane scattering periodicity is retained. However, this overlapping feature alone does not establish the phase assignment; the orthorhombic interpretation is supported collectively by the weak and split reflections observed in Fig. 2 and Figs. S1–S17. However, the peak shape and intensity distribution vary noticeably with deposition method. The Spin thin-film shows a relatively sharp and well-defined (121/200/002) feature with an asymmetric tail, suggesting a more distinct preferred contribution along $q_z$. The DF thin-film also exhibits a clear (121/200/002) peak, but with a broader profile and additional shoulder-like intensity around the main maximum, indicating a less uniform out-of-plane distribution compared with the Spin thin-film. In the DS thin-film, the (121/200/002) region becomes broader, and the intensity gradually increases toward higher $q_z$, suggesting a larger distribution of scattering contributions and a less sharply defined out-of-plane peak response. For the 2 ML DF and DS thin films, the presence of two arcs associated with the (200/002) reflections, located near ($q_r$=1.522 Å$^{-1}$) and ($q_z$=1.522 Å$^{-1}$), indicates a predominantly face-on orientation together with a measurable edge-on contribution (Figs. S1 and S2). By contrast, the absence of the $q_z$ component of this reflection in the 2 ML Spin thin-film is consistent with the higher face-on fraction obtained from the mosaicity analysis (Fig. S3a,e). The formation of the superlattice was further assessed for the 2 ML system from the macroscopic superlattice peaks in the 0.1–0.9 Å$^{-1}$ range. Figure 4d–f present the low-$q$ scattering region for the same thin films, where a series of sharp and regularly spaced peaks are observed. These peaks correspond to harmonic reflections of the nanoplatelet superlattice stacking, and the assigned harmonic orders are shown in Fig. S19. Because the first-order feature lies close to the masked low-$q$ boundary, the fundamental periodicity $q_o$ was obtained by fitting the complete harmonic sequence $q_n \approx nq_o$, rather than by using the first visible maximum alone. The fitted values were approximately 0.1435 Å$^{-1}$ for DF, 0.1395 Å$^{-1}$ for DS, and 0.1466 Å$^{-1}$ for Spin, corresponding, through $d_{SL} = 2\pi/q_o$, to repeat distances of approximately 4.38, 4.50, and 4.29 nm, respectively, which indicate a small expansion of the superlattice period for the DS thin-film. The calculated spacing represents the macroscopic repeating distance of the stacked nanoplatelets, including the inorganic 2 ML platelet thickness together with the outer ligand/shell spacing on both sides. Overall, these results show that whereas all 2 ML thin films form ordered superlattice stacks, the deposition pathway slightly modifies both the out-of-plane scattering profile and the average interparticle stacking distance. The harmonic assignment, fitting algorithm, and corresponding code used to determine the superlattice periodicity are described in the SI and shown in Fig. S19.

Previous work has shown that the maximum thickness of a full double oleic acid ligand layer can be around 3.5-4.5 nm, depending on the compression of the ligand layers[45]. The values obtained in this work lie slightly under these values when subtracting the thickness of the inorganic layers. This can mean that the ligand coverage is not fully dense, which allows for some ligands to lie flat against the surface of the platelets.

The orientational behavior observed in the present work can be rationalized as the result of a competition between particle geometry, ligand-mediated interparticle interactions, and deposition kinetics. Consistent with this interplay, Lee et al. demonstrated that changing the ligand structure modifies both inter-ligand interactions and NPL lateral dimensions, thereby switching $CsPbBr_3$ NPL superlattices between face-on and edge-on orientations[24]. The exceptionally strong face-on orientation of the 2 ML nanoplatelets, irrespective of deposition strategy, is most plausibly attributed to their extreme aspect ratio and unusually large lateral dimensions. With lateral sizes reaching 15-45 nm (Figs. S1j and S2h), these ultrathin nanoplatelets maximize substrate contact when lying parallel to the surface, creating a strong energetic preference for face-on alignment, particularly when the aspect ratio of the lateral dimensions to the thickness is very high. Similar behavior has been reported for anisotropic semiconductor nanoplatelets, where large planar facets promote ordered assembly and orientational anisotropy[46,47]. The fact that the face-on fraction remains dominant even under both fast and slow drop-casting conditions indicates that geometric anisotropy is the primary driving force in the 2 ML regime, overwhelming the comparatively modest changes in assembly kinetics introduced by altering the evaporation rate of the same solvent. This behavior differs from systems in which orientation can be switched solely by changing solvent evaporation dynamics. For example, Ye et al.[22] demonstrated that $CsPbI_3$ nanoplatelet orientation can be tuned between face-on and edge-on configurations by varying solvent vapor pressure during spin coating, whereas Momper et al.[48] showed that the orientation of CdSe nanoplatelets assembled at liquid interfaces can be kinetically controlled through evaporation rate. In the present system, however, changes in deposition and evaporation kinetics were insufficient to overcome the strong face-on preference of the 2 ML NPLs. Because the solvent and ligand chemistry were maintained across the different deposition methods, the observed persistence of face-on orientation indicates that particle geometry, rather than variations in interparticle chemistry, was the dominant factor governing the assembly of the 2 ML NPLs.

The transition from highly oriented 2 ML NPLs to the considerably more disordered 3–5 ML NPL thin films suggests that increasing thickness reduces the dominance of substrate-induced alignment. As the platelet thickness increases, the aspect ratio decreases, and the relative contribution of edge facets, rotational freedom, and ligand-shell-mediated interactions becomes increasingly important. Consequently, the system can access a larger number of orientational configurations, resulting in mixed face-on, edge-on, and unoriented populations. The consistently higher face-on fraction observed in Spin thin-film is therefore likely associated with rapid film thinning and substrate confinement, which suppress extensive three-dimensional reorganization and favor configurations that maximize contact with the substrate before large aggregates can form.

In the characterization of colloidal nanocrystal and quasi-two-dimensional thin films, a fundamental crystallographic distinction must be drawn between short-range local texturing and macroscopic global orientation. Colloidal self-assembly is inherently a multiscale optimization process governed by a delicate interplay of steric, van der Waals, and entropic packing forces mediated by the surface organic ligand shell[49,50]. When these anisotropic building blocks are deposited onto a solid substrate, they can organize into localized mesoscopic clusters or grains that possess a uniform interparticle periodicity[51]. In small-angle X-ray scattering, the strict mathematical repetition within these individual nanoscale domains acts as a coherent interferometer, generating intense fundamental superstructure reflections and clean, high-order harmonics[52]. Critically, however, the presence of these sharp structural features does not by itself dictate macroscopic texturing across the entire thin-film. If these locally ordered domains are tilted, twisted, or randomly oriented relative to one another, they can collectively behave as a powder-like ensemble. Consequently, the MF derived from azimuthal GID line shapes may reveal substantial macroscopic orientational disorder even when coherent interparticle stacking is preserved within the individual domains[43,52]. This distinction explains why the 2 ML samples can exhibit pronounced higher order superlattice coherence while still requiring independent orientational analysis through the MF.

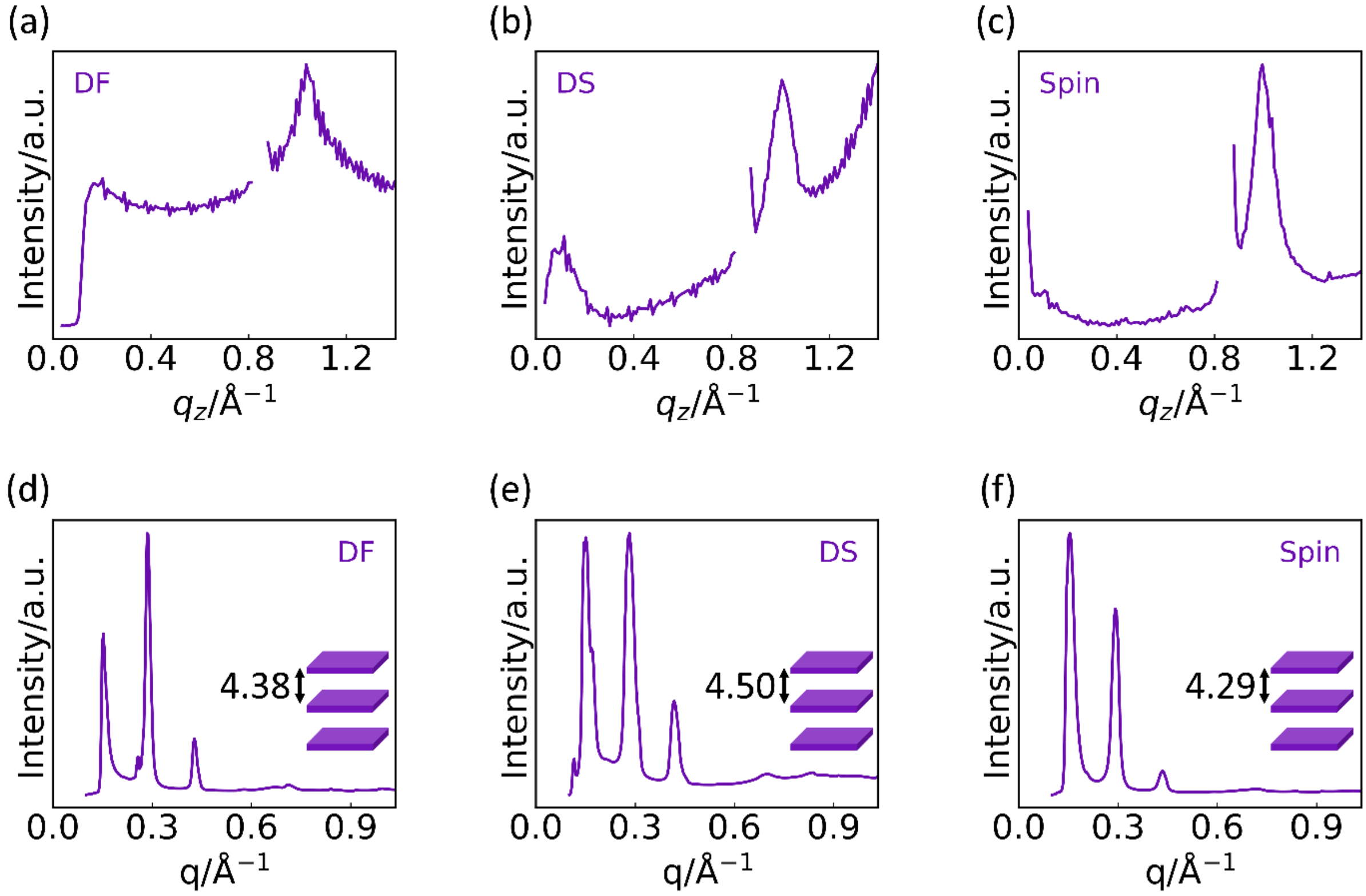

**Figure 4:** Structural analysis of the 2 ML nanoplatelet thin films. (a–c) Integration of the (121/200/002) reflection within the *q*-range of 1.48–1.52 Å$^{-1}$, plotted along the $q_z$ direction to compare the out-of-plane peak behavior for the DF, DS, and Spin thin films. (d–f) Full-ring radial integration of the low-*q* region, showing the superlattice harmonic peaks for the same deposition conditions. The fitted fundamental reciprocal-space periodicity, $q_o$, varies between 0.139 and 0.146 Å$^{-1}$ for the DF, DS, and Spin thin films. The inset shows the superlattice spacing in nm calculated from the periodic harmonics of $q_o$. The harmonic assignment, fitting algorithm, and corresponding code are described in the SI and shown in Fig. S19.

Superlattice fringes require not only preferred orientation but also exceptionally low stacking disorder and a narrow distribution of repeat distances. The combination of extreme aspect ratio, dominant face-on orientation, and uniform platelet thickness therefore creates ideal conditions for coherent multilayer stacking, analogous to the highly ordered nanoplatelet superlattices reported for both perovskite and non-perovskite systems[22,52,53]. At the same time, the slightly larger fitted superlattice period under slower drying is consistent with a modest increase in the average interparticle stacking distance without altering the underlying nanoplatelet structure. The nanocrystals occupy a distinct regime from the NPLs. Under spin-coating and fast drop-casting conditions, the more isotropic NC shape still produces a surprisingly well-defined orientational preference, consistent with facet-driven assembly reported for cuboidal perovskite superlattices. Under slow drop-casting, however, the NC thin films exhibit a larger unoriented fraction. The longer drying time gives the nanocrystals more opportunity to rotate and adopt different configurations during assembly. Because their relatively isotropic shape provides only a weak preference for a particular orientation, this increased mobility leads to a broader distribution of orientations. This contrasts with the 2 ML NPLs, whose extreme aspect ratio strongly favors face-on alignment even during slow drying.

The optical response also reflects deposition-dependent differences in nanoplatelet assembly. The systematically narrower and blue-shifted PL observed for the Spin thin-film, particularly in the 2 ML regime, is consistent with a more uniform distribution of emissive environments and reduced formation of heterogeneous low-energy aggregates. Although the Spin thin-film exhibits the smallest interparticle spacing and the highest degree of orientational order, these characteristics do not necessarily lead to stronger exciton funneling toward lower-energy states. Such funneling also depends on the size and energetic heterogeneity of the assembled domains, as well as on the availability of lower-energy acceptor sites. In contrast, the longer drying time during drop casting allows more extensive nanoplatelet reorganization and the formation of larger stacked or aggregated domains with a broader distribution of local environments. These heterogeneous domains can promote exciton migration toward lower-energy emissive sites, resulting in broader and red-shifted PL. Similar effects have been reported in self-assembled CdSe nanoplatelet stacks, where interparticle interactions and energy transfer alter both the spectral position and exciton dynamics[54,55]. The diminishing magnitude of the spectral shifts with increasing monolayer number and in the nanocrystals further supports an origin related to confinement and assembly rather than crystal structure. As the nanoplatelets become thicker, quantum and dielectric confinement weaken, reducing the sensitivity of the excitonic states to local packing and dielectric environment. Consequently, deposition-induced differences in assembly produce progressively smaller optical changes. The emergence of a shoulder in the 5 ML Spin thin-film can therefore be interpreted as evidence of assembly-related energetic heterogeneity or the coexistence of distinct emissive subpopulations rather than the appearance of a new crystallographic phase, particularly given the persistence of orthorhombic diffraction signatures across all samples.

## Conclusions

The central finding of this work is that $CsPbBr_3$ NPL thin films respond to morphology and deposition hierarchically: the internal orthorhombic-compatible lattice and the 020/040–101/202 lattice–morphology relationship remain largely preserved, whereas superlattice organization and global film texture are strongly modified by NPL thickness and deposition pathway. By comparing 2–5 ML NPLs and larger nanocrystals prepared by slow solvent-evaporation drop-casting, fast solvent-evaporation drop-casting, and spin coating using hexane as a solvent for all deposition methods, we compared the influence of monolayer thickness and deposition pathway while maintaining the same solvent and nominal ligand system. The structural analysis identifies orthorhombic-compatible scattering features across the full NPL series and in the larger NCs, irrespective of deposition method. The clearest evidence is provided by directly resolved splitting near 2.15 $Å^{-1}$ in selected samples and by the weak orthorhombic-compatible reflections observed in the 1.60–2.04 $Å^{-1}$ range. The 2 ML NPLs remain the most challenging case because the asymmetric line shape near 2.15 $Å^{-1}$ does not have a unique structural interpretation. Their assignment therefore relies primarily on the weak orthorhombic-compatible reflections and the broader consistency of the 2D GID dataset, whereas the multi-component fit is treated only as supporting evidence. Within the sensitivity of the present measurements, changing the drying rate or deposition route does not produce a clearly detectable phase transition in the nanoplatelets. The lattice–morphology analysis further shows that this crystallographic relationship is largely preserved across thickness and deposition method. The thickness-related coherence is consistently associated with the 020/040 family of reflections, whereas the lateral crystalline coherence is dominated by the 101/202 reflections. This is consistent with a crystallographic lattice–morphology relationship established during colloidal synthesis and retained during thin-film formation rather than being altered by the deposition process. In contrast to the preserved structural motif, the orientational distribution and superlattice organization are strongly affected by both monolayer number and deposition pathway. The 2 ML NPLs exhibit robust face-on ordering and clear higher-order superlattice harmonics under all deposition conditions, with the stacking periodicity increasing slightly from spin coating to fast and slow solvent-evaporation drop-cast thin films. Thicker NPLs show greater sensitivity to drying kinetics, with slow evaporation enabling greater reorganization and larger edge-on or unoriented populations. Spin coating generally favors face-on alignment, likely because rapid film thinning and substrate confinement limit extensive three-dimensional reorganization. Taken together, these findings establish a multilevel framework for interpreting strongly textured colloidal perovskite films by separating internal lattice structure, the crystallographic lattice–morphology motif, local superlattice coherence, and macroscopic film texture. They reveal a geometry-dominated ultrathin regime, represented by 2 ML NPLs, and a processing-sensitive regime for thicker NPLs. The work resolves previous conflicting results and provides a practical basis for understanding and controlling film texture through nanoplatelet thickness and deposition kinetics and demonstrates why changes in GID intensity or peak visibility should not automatically be interpreted as changes in crystal phase.

## Methods

The preparation of CsPbBr3 NPLs followed the pre-published recipe by Bohn et al.[30] with some modifications.

### Materials

$Cs_2CO_3$ (caesium carbonate, 99%), $PbBr_2$ (lead (II) bromide, 98%), oleic acid (technical grade 90%), oleylamine (technical grade 70%), toluene (for HPLC, 99.9%), acetone (for HPLC, 99.9%), hexane (for HPLC, 97.0%, GC) and were purchased from Sigma-Aldrich.

### Preparation of precursors

Prior to synthesis, oleic acid and oleylamine were individually degassed and purified by heating to 110 °C, followed by filtration through a 0.2 µm PTFE syringe filter to eliminate any insoluble residues. The cesium oleate complex was formulated by dissolving 0.1 mmol of $Cs_2CO_3$ in 10 mL of oleic acid at 100 °C under constant magnetic stirring until a homogeneous solution was obtained. In parallel, the lead bromide precursor was prepared by combining 0.1 mmol of $PbBr_2$ with 100 µL of oleylamine and 100 µL of oleic acid in 10 mL of anhydrous toluene, followed by heating to 100 °C to promote solubilization. The coordinating ligands significantly enhance the solubility of $PbBr_2$ in the apolar medium. Once prepared, both precursor solutions were stored under an inert nitrogen atmosphere in a glovebox to preserve their reactivity and prevent moisture uptake.

### Synthesis of $CsPbBr_3$ NPLs

The colloidal synthesis was conducted entirely at ambient conditions within a standard chemical fume hood. Control over the thickness of the resulting $CsPbBr_3$ NPLs was achieved by modulating the volume ratio between the cesium oleate and lead bromide solutions, as well as by tuning the amount of antisolvent introduced during nucleation. In a typical procedure, a defined volume of Cs-oleate solution was rapidly injected into the $PbBr_2$-ligand solution under vigorous stirring, which is essential to ensure rapid nucleation and uniform crystal growth. After 5 seconds, acetone was added as an antisolvent to facilitate phase separation and induce precipitation of the NPLs. The total stirring time after antisolvent addition was maintained at 1 min. The resulting dispersion was subjected to centrifugation at 4000 rpm for 3 min, and the precipitate was redispersed in anhydrous hexane to yield a stock dispersion with a nominal concentration of 50 mg/mL. To improve colloidal uniformity and eliminate oversized crystallites, a post-synthesis purification step was employed. The hexane dispersion was centrifuged at 3000 rpm for 3 min without the addition of antisolvent. The clear supernatant was carefully extracted using a syringe to minimize perturbation. The concentration of the final colloidal stock was precisely determined gravimetrically and subsequently diluted to 5 mg/mL for standard use in optical and structural measurements, unless otherwise specified. Precursor ratios were optimized to yield NPLs of varying thicknesses. For NPLs comprising 2, 3, 4, and 5 ML, the respective volume ratios of Cs-oleate (µL), $PbBr_2$-ligand solution (mL), and acetone (mL) were as follows: 150/3/2, 150/1.5/2, 150/1.2/2, and 200/1/2. To suppress uncontrolled crystallization in the case of thicker platelets (5 ML), an additional 0.2 mL of acetone was pre-mixed with the $PbBr_2$-ligand solution prior to Cs-oleate injection. For the bigger NCs, i.e., NC1, the ratios were 1000/1/4, whereas an additional 1 mL of acetone was added to the $PbBr_2$-ligand precursor before the Cs-oleate injection. NC2 is the byproduct that is yielded from the 2$^{nd}$ centrifugal process of NC1.

### Thin-film deposition

$SiO_x$ substrates were cleaned by sequential ultrasonication in deionized water, acetone, and isopropanol for 15 min in each solvent. All $CsPbBr_3$ dispersions were adjusted to 50 mg mL$^{-1}$, and 70 µL was deposited onto each substrate. For slow solvent-evaporation drop casting (DS), the dispersion was drop-cast and the Petri dish was immediately sealed to retard hexane evaporation. For fast solvent-evaporation drop casting (DF), the dispersion was drop-cast and left exposed to ambient air to allow rapid evaporation. Spin-coated films were prepared using a two-step programme of 500 rpm for 15 s followed by 1000 rpm for 30 s.

**Characterization**

All characterization experiments were performed at room temperature on thin films obtained by depositing hexane-based NPLs solutions onto silicon substrates, unless otherwise indicated.

PL spectra were acquired using a Renishaw inVia confocal Raman microscope equipped with a 405 nm excitation source (linewidth ≈1 nm).

GID measurements were carried out at the P23 beamline of PETRA III at DESY, Hamburg, using a photon energy of 10 keV. A Pilatus 1M with an active area of 155.1 × 38.4 mm² and a pixel size of 75 × 75 µm$^2$ was used. The sample/SiOx susbtrates were placed in a DCS 500 Anton Paar four-circle goniometer and XYZ stage and tilted to an incidence angle of 0.4°. For each sample, scattering data were averaged over five frames, with an exposure time of 10 s per frame. The scattering data of each frame were normalized and corrected for the transmitted flux. Considering the incidence angle, no other corrections were required[26]. Calibration of the detector images was performed using $LaB_6$ and processed using pyFAI routines. All other data processing, including masking, correction/normalization, 1D integration, 2D maps, 1D and 2D azimuthal data, and fitting, were performed using codes and software developed by the authors[56]. In GID experiments on 2D nanocrystal systems, relying strictly on detector-space coordinates to define crystallite orientation can lead to misleading interpretations of the 2D diffraction data[57]. Although the detector captures all scattered intensity on the physical pixels, the analysis of the scattering process is done in reciprocal space[58]. Therefore, it is essential to distinguish between detector-space and reciprocal-space coordinates for accurate orientation analysis. The pixel coordinates were mapped to $q$ and $\chi$, where $q$ represents the absolute scattering vector magnitude derived from the reciprocal space components (in-plane: $q_x$-$q_y$, out-of-plane: $q_z$), and $\chi$ is the azimuthal angle in detector space[59]. However, because $\chi$ is fundamentally a detector-space angle rather than a pure sample-orientation angle, using it directly to quantify in-plane versus out-of-plane orientation can introduce significant errors, as illustrated in Fig. 2f[57,60]. At low $\chi$ values, the scattering vector lies nearly parallel to the substrate surface and can provide a reasonable reflection of in-plane scattering. At high $\chi$ values, the curvature of the Ewald sphere causes in-plane and out-of-plane scattering components to mix. This geometric mixing makes decoupling the two orientation components challenging without proper reciprocal-space remapping.[57,60] More details about the two approaches are in the SI (Fig. S20).

Consequently, the orientation analysis was performed at two complementary levels. First, the overall macroscopic texture of each thin-film was quantified using the MF following the approach of Ogle et al.[43] . The MF was calculated from the azimuthal intensity distribution by combining a linear angular scaling function centred at the preferred orientation with the normalized, intensity-weighted amplitude at each angular position and summing the resulting contributions over the relevant angular range. For intensity values exactly at 0° or 90°, the weighted amplitude was halved to account for the symmetric contribution extending beyond the 0-90° interval. Importantly, whereas the relative orientation fractions describe how much of the total orientation distribution is associated with each orientation population, the MF describes the degree of alignment within that population; thus, a minor orientation population can nevertheless exhibit a high MF if it is narrowly distributed around its preferred orientation. Second, the orientation of selected crystallographic reflections was examined by analyzing their intensity distribution along the $q_z$ direction. For a given (*hkl*) reflection, intensity appearing at low $q_z$ corresponds predominantly to an in-plane-oriented scattering-vector component, whereas intensity at high $q_z$ corresponds predominantly to an out-of-plane-oriented component. This classification is based on the $q_z$ position of the reflection and is therefore independent of its azimuthal direction within the $q_x$-$q_y$ plane. By combining these two approaches, the global thin-film texture can be distinguished from the orientation of specific crystallographic reflections, allowing structural phase assignment and texture analysis to be treated separately. This distinction is essential for the interpretation of the GID data presented in this study.

High-angular annular dark-field (HAADF) scanning transmission electron microscopy (STEM) images were obtained using an aberration-corrected FEI/Thermo Fisher Titan Themis 200. The microscope used an X-FEG Schottky field emission source operated at an accelerating voltage of 200 kV, a CEOS DCOR+ probe spherical aberration (Cs) corrector, a SuperX Generation 2 energy-dispersive X-ray spectroscopy (EDS) system with four silicon drift detectors (SDDs), a HAADF detector, and an annular dark-field (ADF) detector, besides a 4k × 4k CETA CMOS camera used for conventional TEM imaging. Particle dimensions were measured from TEM images in ImageJ using line analysis. For 2 ML NPLs, both lateral dimensions ($L_1$ and $L_2$) were measured for face-on particles, whereas the platelet thickness (T) was obtained from the smaller edge-on population. A representative lateral dimension for the face-on population was taken as the area-equivalent dimension, $\sqrt{(L_1L_2)}$, and the lateral-to-thickness aspect ratio was estimated as $\sqrt{(L_1L_2)}/T$. For edge-on 2 ML NPLs, the thickness and the accessible lateral dimension (L) were measured directly, and the corresponding aspect ratio was estimated as L/T. For 5 ML NPLs, the thickness and one accessible lateral dimension were measured from edge-on particles. Because the lateral dimensions were approximately equivalent in the TEM images, the accessible lateral dimension was taken as representative of the in-plane size, giving an estimated aspect ratio of L/T. For NC1, the similarity of the relevant lattice spacings prevented an unambiguous assignment of a unique thickness direction. The more isotropic particles were therefore characterized by a single representative edge dimension, whereas the less isotropic particles were characterized by their long and short axes. For NC1, particle anisotropy was expressed as the long-to-short axis ratio rather than as the platelet lateral-to-thickness aspect ratio used for the NPLs. Because some lateral and thickness dimensions were obtained from different orientation subsets, the reported NPL aspect ratios should be regarded as population-level estimates rather than particle-by-particle aspect-ratio distributions.

## Supporting Information

Supplementary information is available in the online version, including additional supplementary figures and text, codes, and experimental details.

## Acknowledgments

We thank the Swedish Energy Agency (P2020-90215) and the Swedish Research Council (Grants no 2023-05244 and 2022-03168) for financial support. GAC and GSA acknowledge funding from the European Union's Horizon Europe research and innovation programme under the Marie Skłodowska-Curie grant agreement No. 101081419 (PRISMAS). UC thanks the partial support by the Wallenberg Initiative Materials Science for Sustainability (WISE) for partial support, funded by the Knut and Alice Wallenberg Foundation. We acknowledge DESY (Hamburg, Germany), a member of the Helmholtz Association HGF, for the provision of experimental facilities. Parts of this research were carried out at PETRA III. Data was collected using beamline P23 operated/provided by DESY Photon Science. We would like to thank Azat Khadiev and Dmitri Novikov for assistance during the experiments. Beamtime was allocated for proposal R-20240724 EC. We acknowledge Myfab Uppsala for providing facilities. Myfab is funded by the Swedish Research Council (2019-00207) as a national research infrastructure.

## Reference

1. Akkerman, Q. A. *et al.* Solution Synthesis Approach to Colloidal Cesium Lead Halide Perovskite Nanoplatelets with Monolayer-Level Thickness Control. *J. Am. Chem. Soc.* **138**, 1010–1016 (2016).
2. Butkus, J. *et al.* The Evolution of Quantum Confinement in CsPbBr3 Perovskite Nanocrystals. *Chem. Mater.* **29**, 3644–3652 (2017).
3. Colloidal Metal-Halide Perovskite Nanoplatelets: Thickness-Controlled Synthesis, Properties, and Application in Light-Emitting Diodes - Otero-Martínez - 2022 - Advanced Materials - Wiley Online Library. https://advanced.onlinelibrary.wiley.com/doi/full/10.1002/adma.202107105.
4. Weidman, M. C., Goodman, A. J. & Tisdale, W. A. Colloidal Halide Perovskite Nanoplatelets: An Exciting New Class of Semiconductor Nanomaterials. *Chem. Mater.* **29**, 5019–5030 (2017).
5. Bertolotti, F. *et al.* Crystal Structure, Morphology, and Surface Termination of Cyan-Emissive, Six-Monolayers-Thick CsPbBr3 Nanoplatelets from X-ray Total Scattering. *ACS Nano* **13**, 14294–14307 (2019).

6. Toso, S., Baranov, D., Giannini, C. & Manna, L. Structure and Surface Passivation of Ultrathin Cesium Lead Halide Nanoplatelets Revealed by Multilayer Diffraction. *ACS Nano* **15**, 20341–20352 (2021).
7. State of the Art and Prospects for Halide Perovskite Nanocrystals | ACS Nano. https://pubs.acs.org/doi/10.1021/acsnano.0c08903.
8. Aboulsaad, M. M., Donzel-Gargand, O., Araujo, R. B. & Edvinsson, T. Symmetry-Driven Phonon Confinement in 2D Halide Perovskites. *J. Phys. Chem. Lett.* **17**, 7008–7019 (2026).
9. Wang, L. *et al.* Ultra-stable CsPbBr3 Perovskite Nanosheets for X-Ray Imaging Screen. *Nano-Micro Lett.* **11**, 52 (2019).
10. Kirschner, M. S. *et al.* Photoinduced, reversible phase transitions in all-inorganic perovskite nanocrystals. *Nat Commun* **10**, 504 (2019).
11. Zhu, H. *et al.* Synthesis of Zwitterionic CsPbBr3 Nanocrystals with Controlled Anisotropy using Surface-Selective Ligand Pairs. *Advanced Materials* **35**, 2304069 (2023).
12. Zhai, W. *et al.* Solvothermal Synthesis of Ultrathin Cesium Lead Halide Perovskite Nanoplatelets with Tunable Lateral Sizes and Their Reversible Transformation into $Cs_4 PbBr_6$ Nanocrystals. *Chem. Mater.* **30**, 3714–3721 (2018).
13. Shamsi, J. *et al.* Colloidal Synthesis of Quantum Confined Single Crystal $CsPbBr_3$ Nanosheets with Lateral Size Control up to the Micrometer Range. *J. Am. Chem. Soc.* **138**, 7240–7243 (2016).
14. Bekenstein, Y., Koscher, B. A., Eaton, S. W., Yang, P. & Alivisatos, A. P. Highly Luminescent Colloidal Nanoplates of Perovskite Cesium Lead Halide and Their Oriented Assemblies. *J. Am. Chem. Soc.* **137**, 16008–16011 (2015).
15. Chen, S. & Gao, P. Challenges, myths, and opportunities of electron microscopy on halide perovskites. *J. Appl. Phys.* **128**, 010901 (2020).
16. Brennan, M. C., Kuno, M. & Rouvimov, S. Crystal Structure of Individual CsPbBr3 Perovskite Nanocubes. *Inorg. Chem.* **58**, 1555–1560 (2019).
17. Diroll, B. T., Banerjee, P. & Shevchenko, E. V. Optical anisotropy of CsPbBr3 perovskite nanoplatelets. *Nano Convergence* **10**, 18 (2023).
18. Hooper, T. J. N., Fang, Y., Brown, A. A. M., Pu, S. H. & White, T. J. Structure and surface properties of size-tuneable CsPbBr3 nanocrystals. *Nanoscale* **13**, 15770–15780 (2021).
19. Wang, Y.-K. *et al.* Chelating-agent-assisted control of CsPbBr3 quantum well growth enables stable blue perovskite emitters. *Nat Commun* **11**, 3674 (2020).
20. Chen, J. *et al.* Oriented Halide Perovskite Nanostructures and Thin Films for Optoelectronics. *Chem. Rev.* **121**, 12112–12180 (2021).
21. Cui, J. *et al.* Efficient light-emitting diodes based on oriented perovskite nanoplatelets. *Science Advances* **7**, eabg8458 (2021).
22. Ye, J. *et al.* Direct linearly polarized electroluminescence from perovskite nanoplatelet superlattices. *Nat. Photon.* **18**, 586–594 (2024).
23. Controlled Assembly and Anomalous Thermal Expansion of Ultrathin Cesium Lead Bromide Nanoplatelets | Nano Letters. https://pubs.acs.org/doi/full/10.1021/acs.nanolett.2c04526.
24. Lee, D. *et al.* Tuning the orientation of self-assembled CsPbBr3 perovskite nanoplatelet superlattices through ligand engineering. *Matter* **9**, 102760 (2026).
25. Govardhan, A. *et al.* From Spin Coating to Zone Casting: Process Dependence and Challenges in Scaling Quasi-2D Perovskite Films. *Mater. Adv.* 10.1039.D6MA00389C (2026) doi:10.1039/D6MA00389C.
26. How to GIWAXS: Grazing Incidence Wide Angle X-Ray Scattering Applied to Metal Halide Perovskite Thin Films - Steele - 2023 - Advanced Energy Materials - Wiley Online Library. https://advanced.onlinelibrary.wiley.com/doi/10.1002/aenm.202300760.
27. Protesescu, L. *et al.* Nanocrystals of Cesium Lead Halide Perovskites (CsPbX3, X = Cl, Br, and I): Novel Optoelectronic Materials Showing Bright Emission with Wide Color Gamut. *Nano Lett.* **15**, 3692–3696 (2015).
28. Wang, Q. *et al.* Quantum confinement effect and exciton binding energy of layered perovskite nanoplatelets. *AIP Advances* **8**, 025108 (2018).
29. Weidman, M. C., Seitz, M., Stranks, S. D. & Tisdale, W. A. Highly Tunable Colloidal Perovskite Nanoplatelets through Variable Cation, Metal, and Halide Composition. *ACS Nano* **10**, 7830–7839 (2016).

30. Bohn, B. J. *et al.* Boosting Tunable Blue Luminescence of Halide Perovskite Nanoplatelets through Postsynthetic Surface Trap Repair. *Nano Lett.* **18**, 5231–5238 (2018).
31. Liu, Y. *et al.* Inkjet-Printed Photodetector Arrays Based on Hybrid Perovskite CH3NH3PbI3 Microwires. *ACS Appl. Mater. Interfaces* **9**, 11662–11668 (2017).
32. Chen, C. *et al.* Efficient Flexible Inorganic Perovskite Light-Emitting Diodes Fabricated with CsPbBr3 Emitters Prepared via Low-Temperature in Situ Dynamic Thermal Crystallization. *Nano Lett.* **20**, 4673–4680 (2020).
33. Wei, M. *et al.* Ultrafast narrowband exciton routing within layered perovskite nanoplatelets enables low-loss luminescent solar concentrators. *Nat Energy* **4**, 197–205 (2019).
34. Aboulsaad, M. M. I. FitED. Zenodo https://doi.org/10.5281/ZENODO.19411620 (2026).
35. Toso, S., Baranov, D., Filippi, U., Giannini, C. & Manna, L. Collective Diffraction Effects in Perovskite Nanocrystal Superlattices. *Acc. Chem. Res.* **56**, 66–76 (2023).
36. Bertolotti, F. *et al.* Coherent Nanotwins and Dynamic Disorder in Cesium Lead Halide Perovskite Nanocrystals. *ACS Nano* **11**, 3819–3831 (2017).
37. Reus, M. A., Reb, L. K., Kosbahn, D. P., Roth, S. V. & Müller-Buschbaum, P. *INSIGHT* : *in situ* heuristic tool for the efficient reduction of grazing-incidence X-ray scattering data. *J Appl Crystallogr* **57**, 509–528 (2024).
38. Wu, T. *et al.* Efficient and Stable $CsPbI_3$ Solar Cells via Regulating Lattice Distortion with Surface Organic Terminal Groups. *Advanced Materials* **31**, 1900605 (2019).
39. Pang, Y. *et al.* Why Do Colloidal Wurtzite Semiconductor Nanoplatelets Have an Atomically Uniform Thickness of Eight Monolayers? *J. Phys. Chem. Lett.* **10**, 3465–3471 (2019).
40. Koster, R. S., Fang, C., van Blaaderen, A., Dijkstra, M. & van Huis, M. A. Acetate ligands determine the crystal structure of CdSe nanoplatelets – a density functional theory study. *Phys. Chem. Chem. Phys.* **18**, 22021–22024 (2016).
41. Dong, Q. *et al.* Electron-hole diffusion lengths > 175 μm in solution-grown $CH_3$ $NH_3$ $PbI_3$ single crystals. *Science* **347**, 967–970 (2015).
42. Chen, A. Z. *et al.* Origin of vertical orientation in two-dimensional metal halide perovskites and its effect on photovoltaic performance. *Nat Commun* **9**, 1336 (2018).
43. Ogle, J., Powell, D., Amerling, E., Smilgies, D.-M. & Whittaker-Brooks, L. Quantifying multiple crystallite orientations and crystal heterogeneities in complex thin film materials. *CrystEngComm* **21**, 5707–5720 (2019).
44. Oshita, N. *et al.* Pure-Blue Emitting Face-Down-Oriented Perovskite Nanoplatelets for Light-Emitting Diodes. *ACS Appl. Nano Mater.* **8**, 8588–8594 (2025).
45. Wetterskog, E. *et al.* Tuning the structure and habit of iron oxide mesocrystals. *Nanoscale* **8**, 15571–15580 (2016).
46. Gao, H. *et al.* Cesium–Lead Bromide Perovskite Nanoribbons with Two-Unit-Cell Thickness and Large Lateral Dimension for Deep-Blue Light Emission. *ACS Appl. Nano Mater.* **3**, 4826–4836 (2020).
47. Scott, R. *et al.* Directed emission of CdSe nanoplatelets originating from strongly anisotropic 2D electronic structure. *Nature Nanotech* **12**, 1155–1160 (2017).
48. Momper, R. *et al.* Kinetic Control over Self-Assembly of Semiconductor Nanoplatelets. *Nano Lett.* **20**, 4102–4110 (2020).
49. Boles, M. A., Engel, M. & Talapin, D. V. Self-Assembly of Colloidal Nanocrystals: From Intricate Structures to Functional Materials. *Chem. Rev.* **116**, 11220–11289 (2016).
50. Bassani, C. L. *et al.* Nanocrystal Assemblies: Current Advances and Open Problems. *ACS Nano* **18**, 14791–14840 (2024).
51. Weidman, M. C., Smilgies, D.-M. & Tisdale, W. A. Kinetics of the self-assembly of nanocrystal superlattices measured by real-time in situ X-ray scattering. *Nature Mater* **15**, 775–781 (2016).
52. Toso, S. *et al.* Multilayer Diffraction Reveals That Colloidal Superlattices Approach the Structural Perfection of Single Crystals. *ACS Nano* **15**, 6243–6256 (2021).
53. Abécassis, B., Tessier, M. D., Davidson, P. & Dubertret, B. Self-Assembly of CdSe Nanoplatelets into Giant Micrometer-Scale Needles Emitting Polarized Light. *Nano Lett.* **14**, 710–715 (2014).
54. Antanovich, A., Prudnikau, A., Matsukovich, A., Achtstein, A. & Artemyev, M. Self-Assembly of CdSe Nanoplatelets into Stacks of Controlled Size Induced by Ligand Exchange. *J. Phys. Chem. C* **120**, 5764–5775 (2016).

55. Gao, Y., Weidman, M. C. & Tisdale, W. A. CdSe Nanoplatelet Films with Controlled Orientation of their Transition Dipole Moment. *Nano Lett.* **17**, 3837–3843 (2017).
56. mustafaaboulsaad - Overview. *GitHub* https://github.com/mustafaaboulsaad.
57. Baker, J. L. *et al.* Quantification of Thin Film Crystallographic Orientation Using X-ray Diffraction with an Area Detector. *Langmuir* **26**, 9146–9151 (2010).
58. Als-Nielsen, J. & McMorrow, D. *Elements of Modern X-Ray Physics*. (Wiley, a John Wiley & Sons, Ltd Publication, Chichester, West Sussex, 2017).
59. Smilgies, D.-M. Scherrer grain-size analysis adapted to grazing-incidence scattering with area detectors. *J Appl Cryst* **42**, 1030–1034 (2009).
60. Rivnay, J., Mannsfeld, S. C. B., Miller, C. E., Salleo, A. & Toney, M. F. Quantitative Determination of Organic Semiconductor Microstructure from the Molecular to Device Scale. *Chem. Rev.* **112**, 5488–5519 (2012).